\documentclass[toc]{PoS}
\usepackage{cite,amsmath}
 \newcommand{\SigmaP}{{\tt Sigma}}
 
 \newcommand{\N}{\nonumber}
 \newcommand{\ep}{\varepsilon}

 \newcommand{\Li}{\mbox{Li}}
 \newcommand{\M}{\mbox{\rm\bf M}}

\title{{\small \rm  DESY 07--111\\  \vspace{-2mm} SFB/CPP-07-41}\\
Difference Equations in Massive Higher 
Order Calculations}

\ShortTitle{Difference Equations in Massive ... \hspace{2.9cm} {\rm I. 
Bierenbaum, J. Bl\"umlein, S.Klein,}
}

\author{I. Bierenbaum, J. Bl\"umlein, S. Klein\\
Deutsches Elektronen--Synchrotron, DESY   \\
Platanenallee 6, D--15738 Zeuthen, Germany
\\
        E-mail: \email{Isabella.Bierenbaum@desy.de, Johannes.Bluemlein@desy.de,
Sebastian.Klein@desy.de}}

\author{\speaker{C. Schneider}
\\
        Research Institute for Symbolic Computation,
        Johannes Kepler University Linz, \\
        Altenberger Str. 69, A-4040, Austria\\
        E-mail: \email{Carsten.Schneider@risc.uni-linz.ac.at}}

\abstract{The calculation of massive 2--loop operator matrix
elements, required for the higher order Wilson coefficients for heavy
flavor production in deeply inelastic scattering, leads to new types
of multiple infinite sums over harmonic sums and related functions,
which depend on the Mellin parameter $N$. We report on the solution
of these sums through higher order difference equations using the
summation package \SigmaP.}

\FullConference{XI International Workshop on Advanced Computing and Analysis Techniques in Physics Research\\
         April 23-27 2007\\
         Amsterdam, the Netherlands}

\begin{document}
\sloppy
%
%
\section{Introduction}
\label{sec:introduction}

\vspace{1mm}\noindent
Single scale quantities in renormalizable quantum field theories, such as
anomalous dimensions and massless Wilson coefficients, are most simply
represented in terms of (finite) harmonic sums.
This representation holds at least up to
3--loop order for massless Yang--Mills theories \cite{MVV1}. The corresponding
Feynman-parameter integrals are such that nested harmonic sums \cite{BK2,summer}
appear in a
natural way, working in Mellin space. They are written either as $S$-- or as $Z$--sums,
\begin{equation}\label{SZSums}
\begin{split}
S_{m_1, \ldots, m_k}(N)& = \sum_{N \geq i_1 \geq i_2 \ldots \geq i_k
> 0} \frac{\prod_{l=1}^k [{\rm
sign}(m_l)]^{i_l}}{i_l^{|m_l|}}\\
Z_{m_1, \ldots, m_k}(N) &= \sum_{N \geq i_1 > i_2 \ldots > i_k > 0}
\frac{\prod_{l=1}^k [{\rm sign}(m_l)]^{i_l}}{i_l^{|m_l|}}~.
\end{split}
\end{equation}
Indeed a wide class of single scale quantities, including the
anomalous dimensions and massless Wilson coefficients for unpolarized and polarized space- and
time-like processes to 3--loop order, the Wilson coefficients for the Drell-Yan process and
pseudoscalar and scalar Higgs boson production in hadron scattering in the heavy quark mass limit,
as well as the soft- and virtual corrections to Bhabha scattering in the on-mass-shell scheme
to 2--loop order \cite{BR} can be represented in this way. Single scale massive quantities at
2 loops, like the unpolarized and polarized heavy-flavor Wilson coefficients in the region $Q^2 \gg
m^2$ -- although the respective Feynman parameter-integrals exhibit a more involved structure
 -- belong to this class too~\cite{HEAV,BBK1,BK1,BBK2}. Finite harmonic sums obey algebraic, cf.
\cite{ALGEBRA}, and structural relations \cite{STRUCT}. The compact representations being
obtained in Mellin space have to be mapped to momentum-fraction space to use the respective
quantities in experimental analyzes. The Mellin inversion requires the analytic continuation of
the harmonic sums~\cite{ANCONT} w.r.t. the Mellin index $N~{\in}~{\bf C}$.

Calculating massive operator matrix elements in Mellin space, which
contribute to the heavy-flavor Wilson coefficients in deeply
inelastic scattering, \cite{BBK1,BBK2,BK1,BBKS1}, new types of
infinite sums occur if compared to massless calculations. In the
latter case, summation algorithms as {\tt summer}~\cite{summer},
{\tt nestedsums}~\cite{nestedsums} and {\tt Xsummer}~\cite{Xsummer}
may be used to calculate the respective sums. {\tt summer} and {\tt
Xsummer} are based on {\tt FORM} \cite{FORM}, while {\tt nestedsums}
is based on {\tt GiNaC}. The new sums which emerged in
\cite{BBK1,BBK2,BK1,BBKS1} can be calculated in different ways. In
Ref.~\cite{BBK1}, we choose analytic methods together with summation.
Another way consists in applying general summation algorithms in the field
of computer algebra. The beginning was made by Gosper's telescoping
algorithm~\cite{GOSPER} for hypergeometric terms\footnote{$f(k)$ is
hypergeometric in $k$ iff $f(k+1)/f(k)=g(k)$ for some fixed rational
function $g(k)$.}. Concerning practical applications, Zeilberger's
extension of Gosper's algorithm to creative
telescoping~\cite{Zeil:91,AequalB} can be considered as the
breakthrough in symbolic summation. The recent summation
package~\SigmaP~\cite{sigma} that can be executed in the computer
algebra system {\tt Mathematica} is going to open up completely new
possibilities in symbolic summation: Based on Karr's
$\Pi\Sigma$-difference fields~\cite{Karr} and further
refinements~\cite{Refined}, the package contains summation
algorithms~\cite{SigmaAlg} that allow to attack not only
hypergeometric sums, like Gosper's and Zeilberger's algorithms, but
sums involving also indefinite nested sums, like, e.g., $S$--sums or
$Z$--sums~\eqref{SZSums}.

In this note we report on using this package for the summation of a
series of new sums and comment on different techniques, which may be
used as well.

\section{The underlying summation principles}

In this section, we discuss methods of the summation package
\SigmaP\ that are relevant to discover and prove the identities
given below. Similar to hypergeometric summation~\cite{AequalB},
\SigmaP\ relies on the following summation techniques.

\noindent {\bf Indefinite summation by telescoping:} {\it Given} an
indefinite sum $S(a)=\sum_{k=0}^af(k)$, {\it find} $g(j)$ such that
\begin{equation}\label{Equ:Tele}
f(j)=g(j+1)-g(j)
\end{equation}
holds within the summation range $0\leq j\leq a$. Then by
telescoping, we get
$$S(a)=g(a+1)-g(0).$$
\noindent{\it Example.} For the summand
\begin{multline}\label{Equ:DoubleSumSummandExp}
f(j)=\frac{(2 j+k+N+2) j! k! (j+k+N)!}{(j+k+1)
   (j+N+1) (j+k+1)! (j+N+1)! (k+N+1)!}\\
+\frac{j! k!
(j+k+N)!\left(-S_1(j)+S_1(j+k)+S_1(j+N)-S_1(j+k+N)\right)}{(j+k+1)!
(j+N+1)!
   (k+N+1)!}
\end{multline}
\SigmaP\ finds
\begin{eqnarray*}
g(j)=\frac{(j+k+1) (j+N+1) j! k! (j+k+N)!
   \Big(S_1(j)-S_1({j+k})-S_1({j+N})+S_1({j+k+N})\Big)}{k N (j+k+1)! (j+N+1)!
   (k+N+1)!}.
\end{eqnarray*}
Therefore summing~\eqref{Equ:Tele} over $j$ produces

\begin{eqnarray}\label{Indef:Exp}
\sum_{j=0}^af(j) &=& \frac{(a+1)! (k-1)! (a+k+N+1)!
   \left(S_1(a)-S_1({a+k})-S_1({a+N})+S_1({a+k+N})\right)}{N (a+k+1)! (a+N+1)! (k+N+1)!}\nonumber\\
& & +\frac{S_1(k)+S_1(N)-S_1({k+N})}{kN(k+N+1)N!} \nonumber\\ & &
+\frac{(2a+k+N+2)a!k!(a+k+N)!}{(a+k+1)(a+N+1)(a+k+1)!(a+N+1)!(k+N+1)!},
\end{eqnarray}
for $a~\in~{\bf N}$.

More generally, given a finite set of indefinite nested sums and
products (in particular, hypergeometric terms and
$Z$--sums/$S$--sums) and given $f(j)$ as a rational expression in
terms of those sums and products, \SigmaP\ decides algorithmically,
whether there exists a rational expression $g(j)$ in terms of those sums
and products such that~\eqref{Equ:Tele} holds; for more details
see~\cite{sigma}.

\bigskip

Usually, one fails to find a solution $g(j)$ for~\eqref{Equ:Tele}.
If $f(j)$, depends on an extra discrete parameter, say $N$,
Zeilberger's {\it creative telescoping paradigm} can be applied.

\medskip

\noindent{\bf Deriving recurrences by creative telescoping:} {\it
Given} an integer $d>0$ and given a sum
\begin{equation}
S(a,N):=\sum_{j=0}^af(N,j)
\end{equation}
with an extra parameter $N$, {\it find} constants
$c_0(N),\dots,c_d(N)$, free of $j$, and $g(N,j)$ such that the
following summand recurrence
\begin{equation}\label{Equ:CreaTele}
c_0(N)f(N,j)+\dots,c_d(N)f(N+d,j)=g(N,j+1)-g(N,j)
\end{equation}
holds for the summation range $0\leq j\leq a$. If one succeeds in
this task, one gets by telescoping the recurrence relation
$$c_0(N)S(a,N)+\dots+c_d(N)S(a,N+d)=g(N,a+1)-g(N,0).$$
Summarizing, one arrives at a recurrence relation of the form
\begin{equation}\label{Equ:Rec}
c_0(N)S(a,N)+\dots c_d(N)S(a,N+d)=q(a,N)
\end{equation}
for the sum $S(a,N)$ of order $d$. Note that
$a$ can be specialized, e.g., to $N$, i.e., one obtains a recurrence
for the definite sum $\sum_{k=0}^{N}f(N,k)$. In other instances, see
below, one can perform the limit $a\to\infty$ which yields a
recurrences for $\sum_{k=0}^{\infty}f(N,k)$.

Creative telescoping has been introduced for hypergeometric
terms~\cite{Zeil:91}. More generally, in \SigmaP\ the summand
$f(N,j)$ may consist of indefinite nested sums and products, in
particular hypergeometric terms and $S$--sums/$Z$--sums.

\medskip

\noindent{\it Example.} For $d=1$ and the summand
\begin{equation}
f(N,j)=\frac{S_1(j)+S_1(N)-S_1({j+N})}{jN(j+N+1)N!}
\end{equation}
 \SigmaP\
computes the solution
\begin{equation}
\begin{split}
c_0(N)&=-N (N+1)^2,\quad c_1(N)(N+1)^3 (N+2),\\
g(N,j)&=\frac{j S(1,j)+(-N-1) S(1,N)-j
   S(1,j+N)-2}{(j+N+1) N!}.
\end{split}
\end{equation}
Hence, summing~\eqref{Equ:CreaTele} over $1\leq j\leq b$ gives
\begin{multline}\label{RecExp}
-N S(N,b)+(1+N)(2+N)S(N+1,b)\\
=\frac{(b+1)\left(S_1(b)+S_1(N)-S_1(b+N)\right)}{(N+1)^2
   (b+N+2) N!}+\frac{b(b+1)}{(N+1)^3 (b+N+1)(b+N+2)N!}
\end{multline}
for the sum $S(N,b)$.

Summarizing, if we succeed in finding a recurrence\footnote{For
simplicity, we suppress extra parameters, like e.g., $a$. Moreover,
we assume that $c_d(N)\neq0$ for all $N\in{\bf N}$.} of type
\begin{equation}\label{Equ:RecWithOuta}
c_0(N)S(N)+\dots c_d(N)S(N+d)=q(N),
\end{equation}
then together with the initial values $S(i)$, $0\leq i<d$, we obtain
an alternative representation of the corresponding input sum,
$\sum_{j=0}^{a}f(N,j)$ (resp.\ $\sum_{j=0}^{N}f(N,j)$ or
$\sum_{j=0}^{\infty}f(N,j)$).

For many applications, such a result is completely satisfactory.
However, if one hunts for a closed form, one can continue as
follows.

\medskip

\noindent{\bf Recurrence solving:} {\it Given} a
recurrence~\eqref{Equ:RecWithOuta}, {\it find} linearly independent
solutions of the homogeneous version of the recurrence, say
$h_1(N),\dots,h_r(N)$, and a particular solution of~\eqref{Equ:Rec},
say $p(N)$.

\smallskip

Namely, if we manage to compute sufficiently many solutions, we can
find constants $k_1,\dots,k_r$ such that
$$S(i)=p(i)+k_1\,h_1(i)+\dots+k_r\,h_r(i)~,$$
for all $0\leq i< d$. As a consequence,
$$S(N)=p(N)+k_1\,h_1(N)+\dots+k_r\,h_r(N),$$
for all $N\geq0$.

With \SigmaP\ we can handle the following situation:
Given~\eqref{Equ:Rec} where the $c_0(N),\dots,c_d(N)$ and $q(N)$ are
given by indefinite nested sums and products, \SigmaP\ can compute
all solutions $S(N)$ in terms of indefinite nested sums and products
(the so-called d'Alembertian solution; see~\cite{AbramovPet}
and~\cite[{Section~7.2.3}]{sigma}).  We emphasize that \SigmaP\
finds, as a special case, all solutions in terms of hypergeometric
expressions and $S$--sums/$Z$--sums.

%
%
\section{\bf\boldmath Summation through difference equations: single sums}
\label{sec:exdiffeq}

\vspace{1mm}\noindent As outlined in Refs.~\cite{BBK1,BK1}, in course
of the calculation of massive 2--loop integrals containing as single
(general) scale the Mellin parameter $N$, different types of sums occur.
These summands are typically products of harmonic sums with
different arguments, weighted by summation parameters and contain
hypergeometric terms, like binomial factors or Beta-function factors
  \begin{eqnarray}
   B(N,i):=\frac{\Gamma(N)\Gamma(i)}{\Gamma(N+i)}. \label{betadef}
  \end{eqnarray}
  Here $i$ is the summation--index, which runs from one to infinity.
  In general, these
  sums can be expressed in terms of nested harmonic sums,
  \cite{BK2,summer} and $\zeta$--values. Note that sums containing
  Beta--functions with different arguments, e.g. $B(i,i),~B(N+i,i)$,
  usually do not lead to harmonic sums in the final result. Further
  quite often binomial sums of the type
  \begin{eqnarray}
   \sum_{i=0}^{N}\binom{N}{i}F(i)~,
  \end{eqnarray}
  with $F(i)$ being a broken--rational function of $i$ times
  a product of harmonic sums,
  emerge. This type of sums can as well be treated with the
  \SigmaP--package but are not considered here.

  Some of these sums can be performed by the existing packages \cite{summer,nestedsums,Xsummer}.
  However, there exists so far no automatic computer program to calculate sums
  which contain Beta--function factors of the type (\ref{betadef})
  and single harmonic sums in the summand.

\subsection{The Sigma-approach}

  As a first example we consider the sum
  \begin{eqnarray}
    T_1(N):=    \sum_{i=1}^{\infty}
                \frac{B(N,i)}{i+N+2}S_1(i)S_1(N+i)~. \label{1Beta25}
  \end{eqnarray}
We treat the upper bound of the sum as a finite integer, i.e., we
consider the truncated version
$$T_1(a,N):=\sum_{i=1}^{a}
                \frac{B(N,i)}{i+N+2}S_1(i)S_1(N+i) ,$$
for $a\in{\bf N}$. Given this sum as input, we execute \SigmaP's
creative telescoping algorithm and find a recurrence for
$T_1(a,N)(=S(a,N))$ of the form~\eqref{Equ:Rec} with order $d=4$.
Finally, we perform the limit $a\to\infty$ and we end up at the
recurrence
\begin{multline*}
-N (N+1)(N+2)^2 \left(4 N^5+68 N^4+455 N^3+1494
   N^2+2402 N+1510\right) T_1(N)\\
-(N+1)(N+2)(N+3) \left(16 N^5+260N^4+1660 N^3+5188 N^2+7912
   N+4699\right) T_1(N+1)\\
+(N+2)(N+4)(2 N+5) \left(4 N^6+74
   N^5+542 N^4+1978 N^3+3680 N^2+3103
   N+767\right) T_1(N+2)\\
+(N+4)(N+5)\left(16 N^6+276
   N^5+1928 N^4+6968 N^3+13716 N^2+13929
   N+5707\right)T_1(N+3)\\
-(N+4)(N+5)^2 (N+6) \left(4 N^5+48 N^4+223
   N^3+497 N^2+527 N+211\right)T_1(N+4)\\
=P_1(N)+P_2(N)S_1(N)\end{multline*} where
\begin{align*}
P_1(N)&=\Big(32 N^{18}+1232 N^{17}+21512
   N^{16}+223472 N^{15}+1514464
   N^{14}+6806114 N^{13}\\
&+18666770N^{12}+15297623 N^{11}-116877645
   N^{10}-641458913 N^9-1826931522N^8\\
&-3507205291 N^7-4825457477 N^6-4839106893 N^5-3535231014
   N^4-1860247616 N^3\\
&-684064448 N^2-160164480 N-17395200\Big) \big/\big(N^3 (N+1)^3
   (N+2)^3 (N+3)^2 (N+4) (N+5)\big)
\end{align*}
and
\begin{align*}
P_2(N)&=-4\Big((4 N^{14}+150 N^{13}+2610
   N^{12}+27717 N^{11}+199197
   N^{10}+1017704 N^9+3786588N^8\\
&+10355813 N^7+20779613 N^6+30225025
   N^5+31132328 N^4+21872237 N^3+9912442N^2\\
&+2672360 N+362400\Big)\big/\big(N^2
   (N+1)^2 (N+2)^2 (N+3) (N+4) (N+5)\big).
\end{align*}
In the next step, we apply \SigmaP's recurrence solver to the
computed recurrence and find the four linearly independent solutions
\begin{align*}
h_1(N)&=\frac{1}{N+2},&h_2(N)&=\frac{(-1)^N}{N(N+1)(N+2)},\\
h_3(N)&=\frac{S_1(N)}{N+2},&
h_4(N)&=\frac{\big(1+(N+1)S_1(N)\big)(-1)^N}{N(N+1)^2(N+2)},
\end{align*}
of the homogeneous version of the recurrence plus the particular
solution
\begin{eqnarray*}
p(N)
     &=&
        \frac{2(-1)^N}{N(N+1)(N+2)}\Biggl[
                                        2S_{-2,1}(N)
                                       -3S_{-3}(N)
                                       -2S_{-2}(N)S_1(N)
                                       -\zeta_2S_1(N)\N\\
&&                                     -\zeta_3
                                       -\frac{2S_{-2}(N)+\zeta_2}{N+1}
                                   \Biggr]
        -2\frac{S_3(N)-\zeta_3}{N+2}
        -\frac{S_2(N)-\zeta_2}{N+2}S_1(N)\N\\
&&      +\frac{2+7N+7N^2+5N^3+N^4}
              {N^3(N+1)^3(N+2)}S_1(N)
          +2\frac{2+7N+9N^2+4N^3+N^4}
              {N^4(N+1)^3(N+2)}
           \end{eqnarray*}
of the recurrence itself. Finally, we look for constants
$c_1,\dots,c_4$ such that
$$T_1(N)=c_1\,h_1(N)+c_2\,h_2(N)+c_3\,h_3(N)++c_4\,h_4(N)+p(N).$$
The calculation of the necessary initial values for $N=0,1,2,3$ does
not pose a problem for \SigmaP\ and we conclude that
$c_1=c_2=c_3=c_4=0$. Hence the final result then reads
\begin{eqnarray}
     T_1(N)
     &=&
        \frac{2(-1)^N}{N(N+1)(N+2)}\Biggl[
                                        2S_{-2,1}(N)
                                       -3S_{-3}(N)
                                       -2S_{-2}(N)S_1(N)
                                       -\zeta_2S_1(N)\N\\
&&                                     -\zeta_3
                                       -\frac{2S_{-2}(N)+\zeta_2}{N+1}
                                   \Biggr]
        -2\frac{S_3(N)-\zeta_3}{N+2}
        -\frac{S_2(N)-\zeta_2}{N+2}S_1(N)\N\\
&&      +\frac{2+7N+7N^2+5N^3+N^4}
              {N^3(N+1)^3(N+2)}S_1(N)
          +2\frac{2+7N+9N^2+4N^3+N^4}
              {N^4(N+1)^3(N+2)}
           ~.\label{2Beta25}
    \end{eqnarray}

Using more refined algorithms of \SigmaP, see e.g.,
\cite{AhlgrenPade}, even a first order difference equation can be
obtained
   \begin{eqnarray}
   &&(N+2)T_1(N)-(N+3)T_1(N+1)\N\\\N\\
   &=&
      2\frac{(-1)^N}{N(N+2)}\Biggl(
                   -\frac{3N+4}
                         {(N+1)(N+2)}\Bigl(\zeta_2+2S_{-2}(N)\Bigr)
                   -2\zeta_3-2S_{-3}(N)-2\zeta_2S_1(N)-4S_{1,-2}(N)
                \Biggr)\N\\
       &&+\frac{S_2(N)-\zeta_2}{N+1}
         +\frac{N^6+8N^5+31N^4+66N^3+88N^2+64N+16}{N^3(N+1)^2(N+2)^3}S_1(N)\N\\
       &&+2\frac{N^5+5N^4+21N^3+38N^2+28N+8}{N^4(N+1)^2(N+2)^2}~.
       \label{Beta25eq2}
   \end{eqnarray}
However, in setting up Eq. (\ref{Beta25eq2}), use had to be made of
further sums of less complexity, which had to be calculated
separately. As above, we can easily solve the recurrence and obtain
again the result~\eqref{2Beta25}.

\medskip

Here and in the following we applied various algebraic relations
between harmonic sums to obtain a simplification of our results,
cf.~\cite{ALGEBRA}.

\subsection{Alternative approaches}

  As a second example we consider the sum
  \begin{eqnarray}
    T_2(N):=\sum_{i=1}^{\infty}\frac{S^2_1(i+N)}{i^2}~,\label{1Harm8}
  \end{eqnarray}
  which does not contain a Beta--function. In a first attempt, we
  proceed as with the first example $T_1(N)$. Namely,
  the naive application of \SigmaP\ yields a fifth order difference
  equation, which is clearly too complex for this sum. However, similar to the
  situation $T_1(N)$, \SigmaP\ can
  reduce it to a third order relation which reads
   \begin{eqnarray}
\label{eq2}
    &&T_2(N)(N+1)^2
    -T_2(N+1)(3N^2+10N+9)\N\\
    &&+T_2(N+2)(3N^2+14N+17)
    -T_2(N+3)(N+3)^2\N\\\N\\
    &=&\frac{
            6N^5+48N^4+143N^3+186N^2+81N-12
            }
            {(N+1)^2(N+2)^3(N+3)^2}
     -2\frac{2N^2+7N+7
            }
           {(N+1)(N+2)^2(N+3)}S_1(N)\N\\
     &&  +\frac{-2N^6-24N^5-116N^4-288N^3-386N^2-264N-72
           }
         {(N+1)^2(N+2)^3(N+3)^2}\zeta_2\label{1diffeqT3}~.
   \end{eqnarray}
Solving this recurrence relation in terms of harmonic sums gives a
closed form; see~\eqref{Harm8} below.

Still  (\ref{eq2}) represents a rather involved way to
solve the problem. A better way consists in first mapping the
numerator $S_1^2(i+N)$ into a linear representation, which can be
achieved using Euler's relation
   \begin{eqnarray}
    S_a^2(N) = 2 S_{a,a}(N) - S_{2a}(N),~~~a > 0~.
   \end{eqnarray}
This is realized in {\tt summer} by the {\sf basis}--command for general--type
harmonic sums,
   \begin{eqnarray}
    T_2(N)=\sum_{i=1}^{\infty}\frac{2S_{1,1}(i+N)-S_2(i+N)}{i^2}
           ~.\label{2Harm8}
   \end{eqnarray}
As outlined in Ref. \cite{summer}, sums of this type can be
   evaluated by considering the difference
   \begin{eqnarray}
    D_2(j)&=&T_2(j)-T_2(j-1)
          =2 \sum_{i=1}^{\infty}\frac{S_1(j+i)}{i}
             -\sum_{i=1}^{\infty}
                                 \frac{1}{i^2(j+i)^2}~.\label{2diffeqT2}
   \end{eqnarray}
   The solution is then obtained by summing the difference (\ref{2diffeqT2}) to
   \begin{eqnarray}
    T_2(N)=\sum_{j=1}^ND_2(j)+T_2(0)~.
   \end{eqnarray}
   The sums in Eq. (\ref{2diffeqT2}) are now calculable trivially
   or are of less complexity than the original sum. In the case considered
   here, only the first sum on the left hand side is not trivial.
   However, after partial fractioning, one can repeat the same procedure, resulting into another
   difference equation, which is now easily calculable.
   Thus using this technique, the solution of Eq. (\ref{1Harm8}) can be obtained
   by summing two first order difference equations or
   solving a second order one. The above procedure is well
   known and some of the summation--algorithms of {\tt summer},
   \cite{summer}, are based on it. As a consequence, infinite sums with an
   arbitrary number of harmonic sums with the same argument
   can be performed using this package. Note that sums containing harmonic
   sums with different arguments, see e.g Eq.~(\ref{1Harm8}, \ref{Harm30}), can in principle
   be summed automatically using the same approach. However,
   this feature is not yet built into {\tt summer}.

  A third way to obtain the sum (\ref{1Harm8}) consists
  of using integral representations of harmonic sums, \cite{BK2}.
  One may represent the sum in terms of the following integrals
  \begin{eqnarray}
   T_2(N)
   &=&2\sum_{i=1}^{\infty}
        \int_0^1dx\frac{x^{i+N}}{i^2}\Bigl(\frac{\ln(1-x)}{1-x}\Bigr)_+
       -\sum_{i=1}^{\infty}
        \Biggl(\int_0^1dx\frac{x^{i+N}}{i^2}\frac{\ln(x)}{1-x}
        +\frac{\zeta_2}{i^2}\Biggr)\N\\
   &=&2\M\Bigl[\Bigl(\frac{\ln(1-x)}{1-x}\Bigr)_+\Li_2(x)\Bigr](N+1)
       -\Biggl(\M\Bigl[\frac{\ln(x)}{1-x}\Li_2(x)\Bigr](N+1)+\zeta_2^2
        \Biggr)~.
       \label{Harm8mellin}
  \end{eqnarray}
  Here the Mellin--transform is defined as
  \begin{eqnarray}
   \M[f(x)](N):=\int_0^1dx~x^{N-1}~f(x)~.
  \end{eqnarray}
  Eq. (\ref{Harm8mellin}) can then be easily calculated since the
  corresponding
  Mellin--transforms are well--known, cf. \cite{BK2}.
  Either of these three methods above leads to
  \begin{eqnarray}
     T_2(N)=
               \frac{17}{10}\zeta_2^2
               +4S_1(N)\zeta_3
               +S^2_1(N)\zeta_2
               -S_2(N)\zeta_2
               -2S_1(N)S_{2,1}(N)
               -S_{2,2}(N)
         ~.
        \label{Harm8}
  \end{eqnarray}

As a third example we would like to evaluate the sum
  \begin{eqnarray}
\label{1Harm27}
     T_3(N) = \sum_{i=1}^{\infty} \frac{S_1^2(i+N) S_1(i)}{i}~.
  \end{eqnarray}
Note that Eq. (\ref{1Harm27})
  is divergent. In order to treat the divergent pieces,
  we introduce the symbol $\sigma_1$
  \begin{eqnarray}
   \sigma_1:=\lim_{a \rightarrow \infty} \sum_{i=1}^{a}\frac{1}{i}~.
\label{sigma1}
  \end{eqnarray}
  The application of \SigmaP\ to the sum (\ref{1Harm27})
  yields a fourth order difference equation
   \begin{eqnarray}
    &&(N+1)^2(N+2)T_2(N)
       -(N+2)\left(4N^2+15N+15\right)T_2(N+1) \N\\
    &&+(2N+5)\left(3N^2+15N+20\right)T_2(N+2)
    -(N+3)\left(4N^2+25N+40\right)T_2(N+3)\N\\
    &&+(N+3)(N+4)^2T_2(N+4)\N\\
    &=&
    \frac{6N^5+73N^4+329N^3+684N^2+645N+215}
         {(N+1)^2(N+2)^2(N+3)^2}
   +\frac{6N^2+19N+9}
          {(N+1)(N+2)(N+3)}S_1(N)\label{1diffeqT2}~,
   \end{eqnarray}
which can be solved.
As in the foregoing example the better way to calculate the sum is to
first change $S_1^2(i+N)$ into a linear basis representation, cf.
\cite{summer},
   \begin{eqnarray}
    T_3(N)=\sum_{i=1}^{\infty}\frac{2S_{1,1}(i+N)-S_2(i+N)}{i}S_1(i)
           ~.\label{2Harm27}
   \end{eqnarray}
One may now calculate $T_3(N)$ using telescoping for the difference
   \begin{eqnarray}
    D_3(j)&=&T_3(j)-T_3(j-1)
          =2\sum_{i=1}^{\infty}\frac{S_{1}(i+j)S_1(i)}{i(i+j)}
-\sum_{i=1}^{\infty}\frac{S_1(i)}{i(i+j)^2}~,\label{2diffeqT3}
   \end{eqnarray}
with
   \begin{eqnarray}
    T_3(N)=\sum_{j=1}^ND_2(j)+T_3(0)~.
   \end{eqnarray}

\noindent One finally obtains
   \begin{eqnarray}
      T_3(N)&=&
                \frac{\sigma_1^4}{4}
                +\frac{43}{20}\zeta_2^2
                +5S_1(N)\zeta_3
                +\frac{3S^2_1(N)-S_2(N)}{2}\zeta_2
                -2S_1(N)S_{2,1}(N) \N\\
              &&+S^2_1(N)S_2(N)
                +S_1(N)S_3(N)
                -\frac{S^2_2(N)}{4}
                +\frac{S^4_1(N)}{4}
        ~.\label{Harm27}
   \end{eqnarray}
\section{A double sum example}

\vspace{1mm}\noindent
Using the Mellin-Barnes integral representation \cite{MB}, massive two--loop integrals
which occur in the calculation of polarized and unpolarized massive operator matrix elements
\cite{BBK1,BBK2,BK1} result into double infinite series of the kind
\begin{eqnarray}
\label{eq1}
S(N)=\sum_{k=0}^{\infty}\sum_{j=0}^{\infty}f(N,k,j)~.
\end{eqnarray}
Here $N~{\in}~{\bf N}$ denotes the Mellin variable. In the following, we consider
an example dealt with in \cite{BBK1,BBK2} where
\begin{align*}
f(N,k,j)&=\frac{\Gamma(k+1)}{\Gamma(k+2+N)}\Gamma(\varepsilon)
\Gamma(1-\varepsilon)\frac{\Gamma(j+1-2\varepsilon)\Gamma(j+1+\varepsilon)
(\Gamma(k+j+1+N)}{\Gamma(j+1-\varepsilon)\Gamma(j+2+N)\Gamma(k+j+2)}\\
&+\frac{\Gamma(k+1)}{\Gamma(k+2+N)}\Gamma(-\varepsilon)\Gamma(1+\varepsilon)
\frac{\Gamma(j+1+2\varepsilon)\Gamma(j+1-\varepsilon)\Gamma(k+j+1+\varepsilon+N)}
{\Gamma(j+1)\Gamma(j+2+\varepsilon+N)\Gamma(k+j+2+\varepsilon)},
\end{align*}
$\varepsilon$ denoting the dimensional regularization parameter of the momentum
integrals performed in $D~=~4~+~\varepsilon$ space-time dimensions. $f(N,k,j)$
obeys the expansion
\begin{equation}
\label{eqSER}
f(N,k,j) = f_0(N,k,j)+\varepsilon
f_1(N,k,j)+\varepsilon^2f_1(N,k,j)+\dots~,
\end{equation}
which is derived using
\begin{equation}
\frac{\Gamma(n+\varepsilon)}{\Gamma(1+\ep) \Gamma(n)} = 1 + \sum_{k=1}^{n-1}
\varepsilon^k Z_{\vec{1}_k}(n-1)~,
\end{equation}
with $\vec{1}_k = \underset{k}{\underbrace{\{1, ..., 1\}}}$,~cf. e.g.~\cite{BK2,MUW}.
Since
the summation could not be accomplished with the available packages
{\tt summer}~\cite{summer}, {\tt nestedsums}~\cite{nestedsums}, or {\tt Xsummer}~\cite{Xsummer},
this double sum was solved in  \cite{BBK1,BBK2} using suitable integral
representations,
which allowed the summation. Here special partial differential operators had to be used
to map the series expansion efficiently. This method clearly was specialized to the
respective cases to be dealt with.

Let us consider the summation of the leading term $f_0(N,k,j)$. The
summation for the higher terms in $\varepsilon$ can be performed in
an analogous way.

We start with the summand $f_0(N,k,j)$ which is given
in~\eqref{Equ:DoubleSumSummandExp}, and find the closed form given
in~\eqref{Indef:Exp}.

Then the limit
\begin{equation}
\label{Equ:Limit1}
\sum_{j=0}^{\infty}f_0(N,k,j)=\lim_{a\to\infty}\sum_{j=0}^af_0(N,k,j)=
\frac{S_1(k)+S_1(N)-S_1(k+N)}{k N (k+N+1)N!}
\end{equation}
is performed leaving us with an expression in finite harmonic sums which depend
on the second summation parameter $k$. We repeat the foregoing procedure and form a finite sum
\begin{equation}
S(N,b):=
\sum_{k=1}^{b}\frac{S_1(k)+S_1(N)-S_1(k+N)}{k N
(k+N+1)N!}\Big(=\sum_{k=1}^{b}\sum_{j=0}^{\infty}f_0(N,k,j)\Big),
\end{equation}
$b~\in~{\bf N}$.

First, we execute \SigmaP's creative telescoping algorithm and
compute the recurrence relation~\eqref{RecExp}. Again, by limit
considerations, it follows that the sum
\begin{equation}\label{Equ:Limit2}
S'(N):=\lim_{b\to\infty}S(N,b)=\sum_{k=1}^{\infty}\sum_{j=0}^{\infty}f_0(N,k,j)
\end{equation}
satisfies the recurrence relation
$$-N S'(N) + (1 + N) (2 + N) S'(N+1)=\frac{(N+1)S_1(N)+1}{(N+1)^3 N!}.$$
Next, we use \SigmaP's recurrence solver and compute the general
solution
$$\frac{1}{N (N+1) N!}c+\frac{S_1(N)^2+S_2(N)}{2 N (N+1)
N!}$$
for a constant $c$. Checking the initial value $S'(1)$ shows
that
$$S'(N)=\frac{S_1(N)^2+S_2(N)}{2 N (N+1)!}.$$
Hence,
$$\sum_{k=0}^{\infty}\sum_{j=0}^{\infty}f_0(N,k,j)=
\frac{S_1(N)^2+S_2(N)}{2 N (N+1)!}+\sum_{j=0}^{\infty}f_0(N,0,j).$$
Using again \SigmaP\ we find
\begin{eqnarray}
\sum_{j=0}^af_0(N,0,j) &=& \sum_{j=0}^a\frac{2 j+N+2}{(j+1)^2
(j+N+1)^2(N+1)!}
\nonumber\\
&=& \frac{2 a+N+2}{(a+1)^2
(a+N+1)^2(N+1)!}
\nonumber\\ & &
+\frac{S_2(a)}{N(N+1)!}+\frac{S_2(N)}{N(N+1)!}
-\frac{S_2(a+N)}{N(N+1)!}.
\end{eqnarray}
In the limit $a\to\infty$ this later expression simplifies to
$$\frac{S_2(N)}{N (N+1)!}~.$$
Finally, we find
$$\sum_{k=0}^{\infty}\sum_{j=0}^{\infty}f_0(N,k,j)=
\frac{S_1(N)^2+3S_2(N)}{2 N (N+1)!}.$$
Completely analogously, we
compute, for instance, the linear term in the series
expansion~\eqref{eqSER}:
$$\sum_{k=0}^{\infty}\sum_{j=0}^{\infty}f_1(N,k,j)=\frac{-S_1(N)^3-3 S_2(N)S_1(N)-8
   S_3(N)}{6 N (N+1)!}.$$

%
%
\begin{appendix}
 \section{Appendix}
 \label{sec:Appendix}
 \renewcommand{\theequation}{\thesection.\arabic{equation}}

\vspace{1mm}\noindent
  Further examples of sums calculable by \SigmaP\ and the order of their
  respective recurrence relation are given below. Note that the inhomogeneous
  part of the recurrence relations are given in terms of single harmonic
  sums up to depth $2$, $\zeta$--values and polynomials in $N$ only.
Below we list a series of examples for infinite sums, which
emerged calculating the linear terms in $\epsilon$ for the massive operator matrix elements
\cite{BBKS1}.
The sums were calculated solving higher order
difference equations with \SigmaP .

  \subsection{\bf\boldmath Sums resulting in fourth order difference equations}
       \begin{eqnarray}
    \sum_{i=1}^{\infty}
    \frac{B(N,i)}{i}S_1(i+N)^2
     &=&
       \frac{17}{10}\zeta_2^2
       +4S_1(N)\zeta_3
       +2S_{-2}(N)\zeta_2
       +S^2_1(N)\zeta_2
       -3S_4(N)\N\\
&&     +2S^2_{-2}(N)
       -4S_1(N)S_3(N)
       -S^2_1(N)S_2(N)
       +\frac{S^2_1(N)}{N^2}
       +2\frac{S_1(N)}{N^3}\N\\
&&     +\frac{2}{N^4}
       -(-1)^N\frac{2S_{-2}(N)+\zeta_2}{N^2}
           ~,\label{Beta32}\\
    \sum_{i=1}^{\infty}\frac{S_1(i)S_2(i+N)}{i}
     &=&
               \frac{\sigma_1^2}{2}\zeta_2
               -\frac{1}{5}\zeta_2^2
               -S_1(N)\zeta_3
               -\frac{S^2_1(N)-3S_2(N)}{2}\zeta_2
               -2S_{3,1}(N)\N\\
&&             +\frac{1}{2}S_4(N)
               +\frac{S^2_1(N)S_2(N)}{2}
               +S_1(N)S_3(N)
~,
         \label{Harm30}\\
    \sum_{i=1}^{\infty}\frac{S_1(i+N)S_2(i+N)}{i}
     &=&
               \frac{\sigma_1^2}{2}\zeta_2
               -\frac{1}{5}\zeta_2^2
               -S_1(N)\zeta_3
               +\frac{1}{2} [S_2(N)-S_1^2(N)] \zeta_2
               +S_{2,2}(N)\N\\
               &&+S^2_1(N)S_2(N)
               +S_1(N)S_3(N)
~.
         \label{Harm29}
   \end{eqnarray}
  \subsection{\bf\boldmath Sums resulting in third order difference equations}
   \begin{eqnarray}
    \sum_{i=1}^{\infty}
    \frac{B(N,i)}{(i+N+1)^2}S_1(i)
     &=&(-1)^N\frac{2S_{1,-2}(N)+S_{-3}(N)+\zeta_2S_1(N)+\zeta_3}{N(N+1)}
        +\frac{\zeta_2-S_2(N)}{(N+1)^2}
           ~,\label{Beta8}\\
   \sum_{i=1}^{\infty}
    \frac{B(N,i)}{(N+i+2)^2}S_1(i)
     &=&
        \frac{(-1)^{N}}{N(N+1)(N+2)}
        \Bigl(
              2\zeta_3
              +4S_{1,-2}(N)
              +2S_{-3}(N)
              +2S_1(N)\zeta_2\N\\
&&            +\frac{2\zeta_2}
                   {N+1}
              +\frac{4S_{-2}(N)}
                    {N+1}
         \Bigr)
              +\frac{\zeta_2-S_2(N)}
                    {(N+2)^2}
              +\frac{3N^2+N-8}
                   {N(N+1)^4(N+2)^2}
                      ~,\label{Beta29}\\
    \sum_{i=1}^{\infty}
    \frac{B(N,i)}{i}S_2(N+i)
     &=&
        \frac{7}{10}\zeta_2^2
        -2S_{-2}(N)\zeta_2
        +S_2(N)\zeta_2
        -3S_4(N)
        -2S_{-2}(N)^2
        -S_2(N)^2
\N\\
&&      +(-1)^N \frac{\zeta(2)+2S_{-2}(N)}{N^2}
        +\frac{S_2(N)}{N^2}
           ~.\label{Beta26}
       \end{eqnarray}
\end{appendix}

\vspace{3mm} \noindent {\bf Acknowledgment}.~ We thank J. Vermaseren
for discussions. This paper was supported in part by
Sonderforschungsbereich Transregio-9, Computergest\"utzte
Theoretische Teilchenphysik, and the Studienstiftung des Deutschen
Volkes. C.~Schneider was supported by the SFB grant F1305 of the
Austrian Science Foundation FWF.

%
%

\end{document}